\documentclass[usenatbib]{mnras}
\usepackage{graphicx}
\usepackage{txfonts}

\usepackage{url}
\usepackage{graphicx}
\usepackage[hyphenbreaks]{breakurl}
\usepackage{float}
\usepackage{subfig}
\usepackage{multirow}
\usepackage{natbib}

\title[Small-scale impulsive events on the active Sun]{Power-law energy distributions of small-scale impulsive events on the active Sun: Results from IRIS}

\author[]
{Nived Vilangot Nhalil,$^{1,2}$ Chris J. Nelson,$^{2}$ Mihalis Mathioudakis,$^{2}$  \newauthor J. Gerry Doyle,$^{1}$
Gavin Ramsay$^{1}$ \\
$^{1}$Armagh Observatory \& Planetarium, College Hill, Armagh, BT61 9DG, UK\\
$^{2}$Astrophysics Research Centre (ARC), School of Mathematics and Physics, Queens University, Belfast, BT7 1NN, Northern Ireland, UK}

\begin{document}
\outer\def\gtae {$\buildrel {\lower3pt\hbox{$>$}} \over 
{\lower2pt\hbox{$\sim$}} $}
\outer\def\ltae {$\buildrel {\lower3pt\hbox{$<$}} \over 
{\lower2pt\hbox{$\sim$}} $}
\newcommand{\Msun}{$M_{\odot}$}
\newcommand{\lsun}{$L_{\odot}$}
\newcommand{\Rsun}{$R_{\odot}$}
\newcommand{\solar}{${\odot}$}
\newcommand{\kep}{\sl Kepler}
\newcommand{\ktwo}{\sl K2}
\newcommand{\tess}{\sl TESS}
\newcommand{\swift}{\it Swift}
\newcommand{\Porb}{P_{\rm orb}}
\newcommand{\nuorb}{\nu_{\rm orb}}
\newcommand{\eplus}{\epsilon_+}
\newcommand{\eminus}{\epsilon_-}
\newcommand{\cd}{{\rm\ c\ d^{-1}}}
\newcommand{\MdotL}{\dot M_{\rm L1}}
\newcommand{\Mdot}{$\dot M$}
\newcommand{\Mdotsolar}{\dot{M_{\odot}} yr$^{-1}$}
\newcommand{\Ldisk}{L_{\rm disk}}
\newcommand{\src}{KIC 9202990}
\newcommand{\ergscm} {erg s$^{-1}$ cm$^{-2}$}
\newcommand{\rchi}{$\chi^{2}_{\nu}$}
\newcommand{\chisq}{$\chi^{2}$}
\newcommand{\pcmsq} {cm$^{-2}$}
\outer\def\gtae {$\buildrel {\lower3pt\hbox{$>$}} \over 
{\lower2pt\hbox{$\sim$}} $}
\outer\def\ltae {$\buildrel {\lower3pt\hbox{$<$}} \over 
{\lower2pt\hbox{$\sim$}} $}

\providecommand{\lum}{\ensuremath{{\cal L}}}
\providecommand{\mg}{\ensuremath{M_{\rm G}}}
\providecommand{\bcg}{\ensuremath{BC_{\rm G}}}
\providecommand{\mbolsun}{\ensuremath{M_{{\rm bol}{\odot}}}}
\providecommand{\teff}{\ensuremath{T_{\rm eff}}}


\maketitle
\begin{abstract}
Numerous studies have analysed inferred power-law distributions between frequency and energy of impulsive events in the outer solar atmosphere in an attempt to understand the predominant energy supply mechanism in the corona. Here, we apply a burst detection algorithm to high-resolution imaging data obtained by the Interface Region Imaging Spectrograph to further investigate the derived power-law index, $\gamma$, of bright impulsive events in the transition region. Applying the algorithm with a constant minimum event lifetime (of either $60$ s or $110$ s) indicated that the target under investigation, such as Plage and Sunspot, has an influence on the observed power-law index. For regions dominated by sunspots, we always find $\gamma <2$; however, for datasets where the target is a plage region, we often find that $\gamma >2$ in the energy range [$\sim10^{23}$, $\sim10^{26}$] erg. Applying the algorithm with a minimum event lifetime of three timesteps indicated that cadence was another important factor, with the highest cadence datasets returning $\gamma >2$ values. The estimated total radiative power obtained for the observed energy distributions is typically 10 -- 25\% of what would be required to sustain the corona indicating that impulsive events in this energy range are not sufficient to solve coronal heating. If we were to extend the power-law distribution down to an energy of $10^{21}$ erg, and assume parity between radiative energy release and the deposition of thermal energy, then such bursts could provide 25 -- 50\% of the required energy to account for the coronal heating problem.

\end{abstract}


\section{Introduction}

How the coronae of the Sun and solar-like stars are heated to multi-million degree temperatures remains an open question in modern astrophysics. Typically, energy supply is explained as coming either from magnetohydrodynamic (MHD) waves (\citealt{Alfven47}) or magnetic reconnection (\citealt{parker1988}). However, because the spatial scales of dissipation from either mechanism would be so small (km length-scales or less) neither has currently been shown to be dominant in the Sun and other solar-type stars. It is well known that large-scale solar flares are magnetic reconnection events which can release huge amounts of energy ($\sim$10$^{32}$ erg) over time-scales of the order of minutes, yet these events are sufficiently rare that their time-averaged energy contribution is not high enough to compensate for the radiative losses of the corona. On sub-arcsecond scales, a host of magnetic reconnection associated features include, but are not limited to, Ellerman bombs (EBs; \citealt{Ellerman1917, Vissers2013, Nelson2015, reid2016}), Quiet-Sun Ellerman-like Brightenings (QSEBs; \citealt{Rouppe2016, Nelson2017}), Explosive Events (EEs; \citealt{Brueckner1983, Huang2017, Huang2018}), and UV bursts (\citealt{Peter2014, Nelson16, Young2018}) have been shown to be common in the solar atmosphere, but the ability of these features to deposit enough energy to heat the upper layers of the Sun remains uncertain.
 
\begin{figure*}
\vspace*{-2.5cm}

\includegraphics[width=1.0\textwidth]{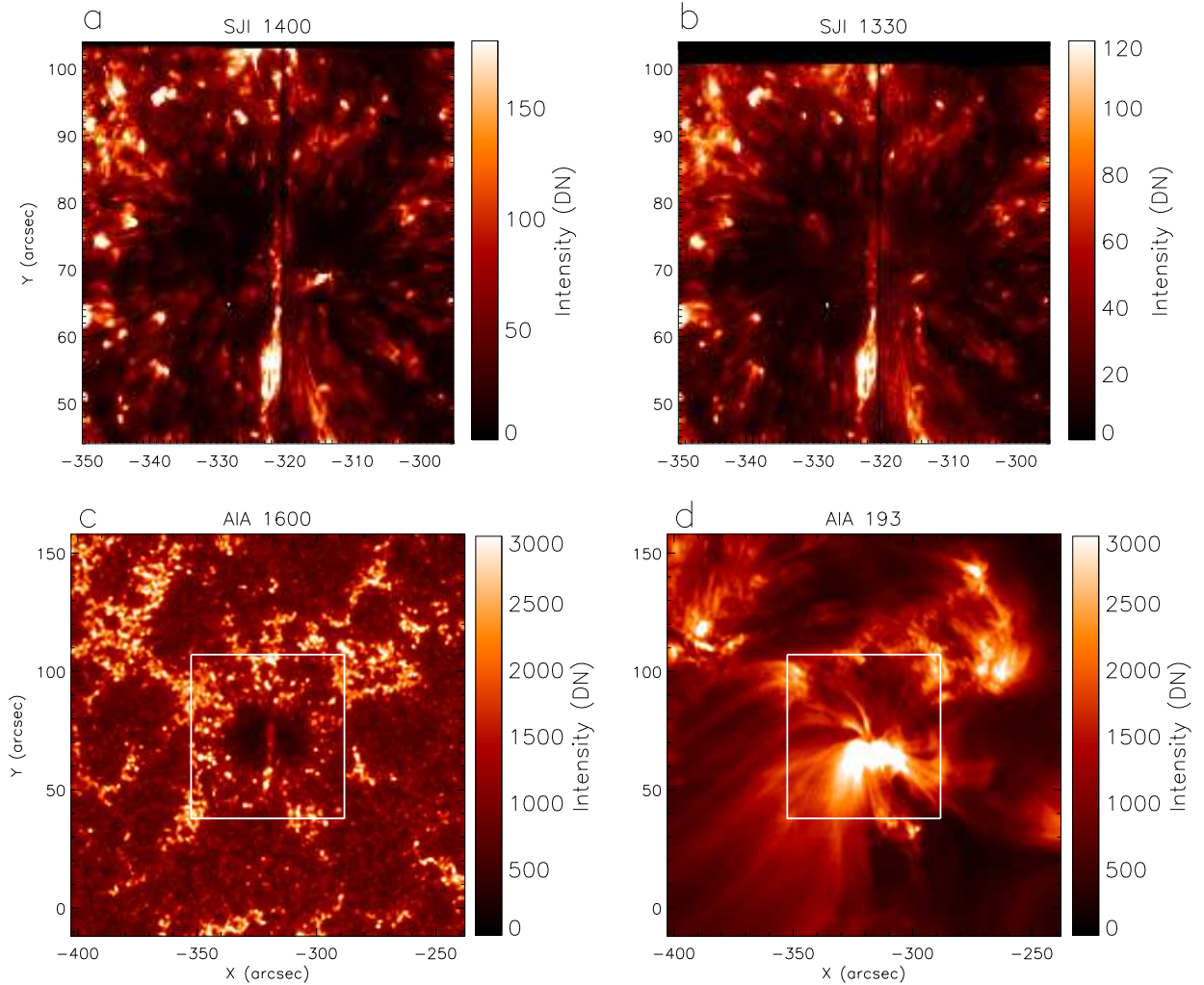}
\vspace*{-2.5cm}

\caption{The studied FOV within AR $11836$ (Dataset 1) sampled on August 31 2013 at approximately 15:59 UT. The SJI 1400 {\AA} and 1300 {\AA} images are plotted in panels (a) and (b), respectively. Context imaging from the SDO/AIA AIA 1600 {\AA} (c) and 193 {\AA} (d) filters are also plotted. The white box over-laid on these panels identifies the FOV of the IRIS SJI observation. }
\label{fig1}
\end{figure*}

 Two numbers are often cited by researchers when discussing the ability of magnetic reconnection events to provide the $10^7 ergs^{-1}cm^{-2}$ required to heat the upper layers of the active solar atmosphere (\citealt{Withbroe77}). The first is that in order to understand whether magnetic reconnection is relevant to coronal heating we must be able to probe  ‘nano-flare’ events with energies less than $\sim$10$^{24}$ erg (\citealt{parker1988}). This work built on \cite{Levine1974} who proposed that coronal heating was due to a multitude of small reconnections. Despite the apparent simplicity of this task, it has so far proved difficult to identify large samples of events at these energies due to the observational stipulations required. Obtaining both the high spatial and temporal resolution needed in the upper layers of the solar atmosphere with a large enough field-of-view (FOV) to provide some general information is still a non-trivial task. The second number often quoted is that in order to explain the heating of the corona by magnetic reconnection events the energy versus frequency distribution must have a power-law with an index, $\gamma$, steeper than 2 (\citealt{Hudson1991}). If $\gamma>$2, magnetic reconnection events with lower energies are important in heating the corona; however, if $\gamma<$2, higher energy flares are dominant and lower energy events such as nano-flares cannot contribute much to the total power \citep{Benz1998,Parnell2000}. Therefore, investigating the frequencies of suitably small impulsive burst events (often assumed to be an observational signature of magnetic reconnection) in the upper atmosphere remains an essential task in solar physics.

\begin{table*}
\caption{Details of the IRIS SJI observations studied here. The 'Cadence (s)' column refers to the IRIS SJI 1400 \AA\ channel cadence. The * indicates that IRIS $1330$ \AA\ SJI data are also available with the same cadence.}
\centering
\begin{tabular}{ |c|c|c|c|c|c|c|c|c|c| }
\hline
Data  & Date of      & Start Time  & End time    & FOV   & ($x_\mathrm{c}$, $y_\mathrm{c}$) & Cadence  &NOAA   & Target   & OBSID \\
      & Observation  & (UT)      &(UT)        & (arcsec)   & (arcsec)     & (s)      &Number  &  &\\
\hline
Set 1*  & 2013/08/31   &15:59   &17:54        & $60''\,\times61\,''$  & (-332'', 74'')       & 12 & 11836 &Sunspot & 4000255147 \\
Set 2*  & 2013/12/17   &00:34   &01:33        & $167''\,\times174\,''$ & (266'', 177'')   &20 & 11921 &Sunspot/Plage & 3820256107 \\
Set 3* & 2015/04/16   &18:54   &03:20+1day   & $59''\,\times60\,''$    & (-122'', 270'')   &19 &12321 & Sunspot/Plage & 3820009270 \\
Set 4* & 2015/11/12   &02:57   &03:57        & $120''\,\times119\,''$ & (-134'', -298'') &13 &12449 &Plage & 3600104017\\
Set 5*  & 2015/11/12   &15:57   &16:50        & $120''\,\times119\,''$ & (63'', -244'') &21 &12449 & Plage & 3600106007\\
Set 6  & 2017/03/26   &11:31   &11:51        & $119''\,\times119\,''$ & (-191'', 252'')   &5 &12643 & Plage & 3613107603 \\
Set 7  & 2017/03/26   &11:02   &11:22        & $119''\,\times119\,''$ & (-191'', 252'')   &3 &12643 & Plage & 3613105603 \\
Set 8  & 2017/05/01   &13:59   &16:59        & $120''\,\times119\,''$ & (-317'', 251'')   &12 &12654 & Plage & 3620104423 \\
Set 9  & 2016/06/06   &13:59   &16:51        & $120''\,\times119\,''$ & (147'', -118'')   &12 &-----& Plage & 3620104423 \\
Set 10  & 2014/01/07   &13:44   &14:38        & $167''\,\times174\,''$ & (-159'', -85'')   &19 &11944 & Sunspot/Plage & 3860259472 \\
Set 11  & 2016/01/14   &18:11   &19:07        & $119''\,\times119\,''$ & (387'', 355'')   &11 &12483 & Plage & 3600257420 \\
Set 12  & 2016/01/19   &15:09   &16:03        & $60''\,\times65\,''$ & (-18'', 313'')   &11 &12485 & Plage & 3610257419 \\

\hline
\end{tabular}

\label{tab1}
\end{table*}

 Numerous authors have attempted to analyse the power-law distributions of brightenings in the upper solar atmosphere using a variety of data. \cite{Shimizu1995}, for example, looked at active region (AR) transient brightenings and found that their frequency distribution as a function of energy had a power-law with an index of 1.64 to 1.89. Interestingly, it was found that the returned power-law index was dependent upon the pixel size with smaller pixels providing larger indices. Notably though, the events studied by \citet{Shimizu1995} had energies in excess of 10$^{27}$ erg, well above the 10$^{24}$ erg level suggested by \citet{parker1988}. \cite{Parnell2000} investigated Transition Region and Coronal Explorer (TRACE; \citealt{Golub1999}) events with energies in the range 10$^{23}$-10$^{26}$ erg deriving a power-law index greater than two. However, those authors suggested that the input from this energy range was insufficient to heat the quiet solar corona, implying that even smaller energies, down to 10$^{21}$ erg, would be required. \cite{Aschwanden2002} expanded on this work and looked at TRACE data from the 171 \AA\ and 195 \AA\ filters deriving power-law indices of 1.86 and 1.81, respectively.  The above dependence on instrument and pixel scale highlights the importance of temporal and spatial resolution in investigating the smallest and faintest events. 

 In almost all observational studies conducted to date, the derived power-law index between frequency and energy has not been sufficient to provide enough energy to the upper atmosphere to compensate for radiative losses. There is, therefore, currently very little confirmatory evidence to support the hypothesis that micro/nano-flaring is sufficient to heat the coronae. Analytical and numerical modelling by a host of authors, (e.g., \citealt{Cargill2004, Rappazzo2008, Parnell2010, Bowness2013, Jess2019, Mondal2020}), offer some hints that further observational studies would be beneficial. 
 Additionally, research on solar-like stars could offer some clues about the contribution of magnetic reconnection to the heating of the upper atmosphere. For example, \cite{Doyle1985} presented a linear correlation between the X-ray flux of quiescent dMe stars (plus the Sun) against the time-averaged energy emitted by flares in a near ultra-violet wave-band.  Various authors have looked at active region mostly when studying nano-flaring (e.g.:  \citealt{Antiochos_2003, Testa_2013, Graham_2019}) finding constraints on parameters such as, the time interval between heating events in a flux tube and the variation in magnitude between events.

Here, we look again at this problem using high spatial and temporal resolution observations obtained by the Interface Region Imaging Spectrograph (IRIS; \citealt{Pontieu2014}). This instrument has an effective spatial resolution of 0.33$''$ allowing us to probe the transition region of the Sun down to sub-arcsecond scales. Our work is set out as follows: In Sect.~\ref{Observations} we outline the data studied here and the detection algorithm employed; In Sect.~\ref{Results} we present the results obtained through applying our algorithm, and in Sect.~\ref{Discussion} we include a brief discussion and draw our conclusions.

\section{Methods}
\label{Observations}

\subsection{Observations}

The aim of this work is to analyse burst events with short lifetimes in the transition region of Active Regions (ARs) meaning high-cadence (\ltae 20 s) in the Slit-Jaw Imager (SJI) 1400 {\AA} channel is essential. Additionally, we require a near constant field-of-view (FOV) in order to analyse the entire lifetimes of bursts implying that data must have been recorded in either sit-and-stare mode (in which the IRIS slit is fixed at a particular location on the Sun and continuously tracks the solar rotation) or in dense raster mode with a maximum of four steps ($\sim1''$ shift in FOV). Searching the IRIS data catalogue with these criteria returned several hundred candidate datasets totaling tens of TB in size. However, as analysing all of these datasets was not possible at this time, here we focus instead on a smaller sample of $12$ datasets in order to understand some general properties of bursts which can be used to infer interesting results and direct future research. The details of these $12$ SJI datasets are presented in Table \ref{tab1}, where the stars indicate the five datasets which also include simultaneous imaging from the IRIS SJI 1330 {\AA} channel.  These data were downloaded as Level-2 IRIS data products, on which dark current removal, flat-field, geometrical distortion, orbital and thermal drift corrections had been applied.  
\begin{figure}
\includegraphics[width=0.5\textwidth]{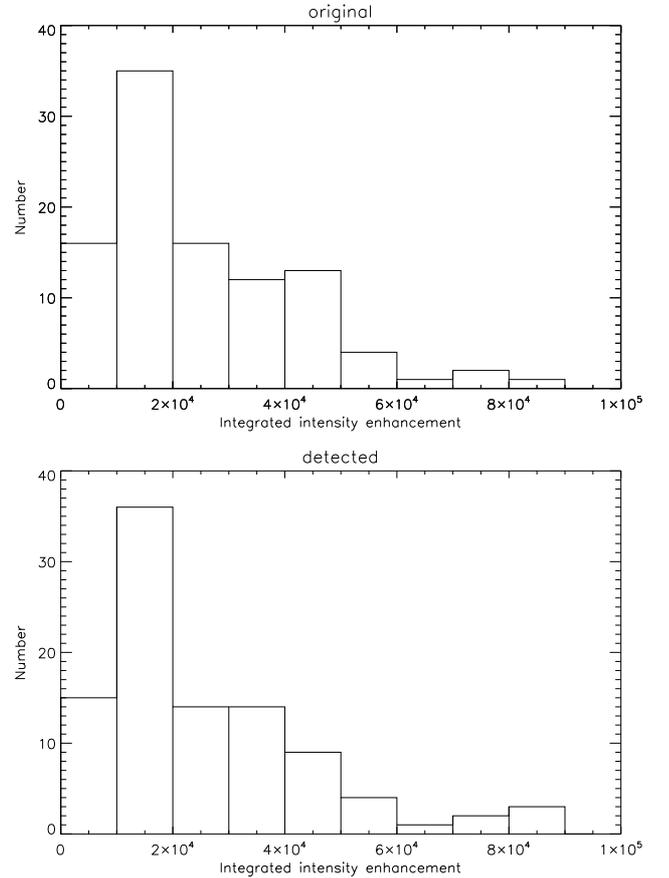}
\caption{Top panel shows the histogram for the actual value of integrated intensity enhancement in the burst. A similar plot is shown in the bottom panel but for bursts detected using 3 $\sigma$ detection condition.}
\label{fig_test}
\end{figure}

\begin{figure*}
\includegraphics[width=1.0\textwidth]{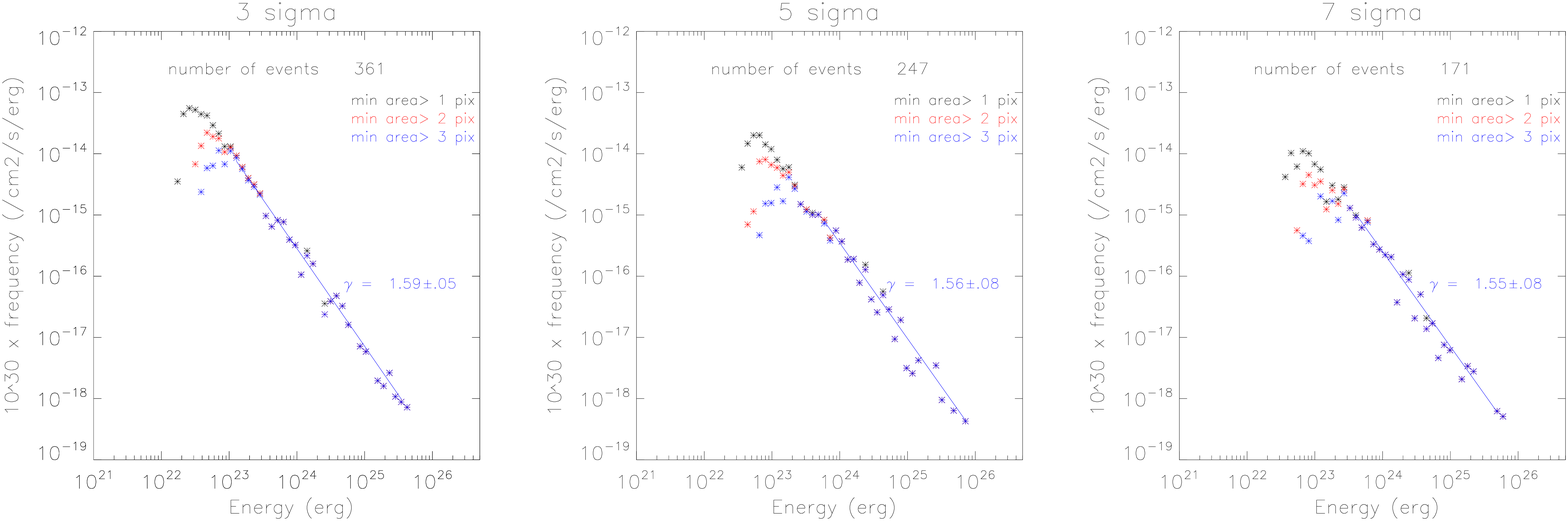}
\includegraphics[width=1.0\textwidth]{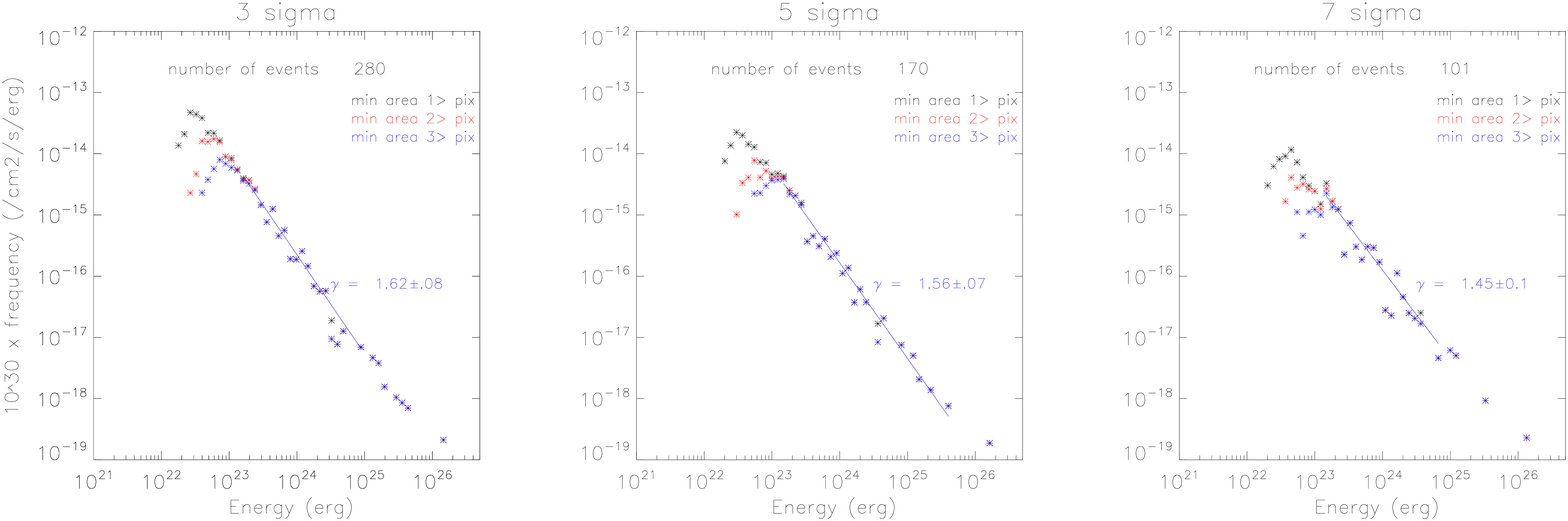}
\caption{The effect of changing the intensity threshold from 3$\sigma$ (left column) to 5$\sigma$ (middle column) to 7$\sigma$ (right column) above the background, for varying minimum areas, on the detection of impulsive bursts in Dataset 1 for the SJI 1400 \AA\ (top row) and $1330$ \AA\ (bottom row) channels. The distribution of bursts with area $>$1, $>$2, $>$3 pixels are shown in black, red, blue respectively. The effects of the minimum area condition are most apparent at low energies (below approximately $10^{23}$ erg). The over-laid 'number of events' indicates the total number of events detected with the area $>$3 pix.}
\label{fig2}
\end{figure*}

The SJI 1400 {\AA} channel is dominated by the Si {\sc iv} 1394 {\AA} and 1403 {\AA} resonance lines which are formed in the transition region, while the SJI 1330 {\AA} channel is dominated by the C {\sc ii} 1335 {\AA} and 1336 {\AA} lines which are formed in the upper chromosphere. The mean formation temperature of C {\sc ii} and Si {\sc iv} lines are $\sim 3\times10^4$ K and $\sim 8\times 10^4$ K respectively \citep{Rathore_1,Rathore_2}. Therefore, these two SJI channels provide an opportunity to study the energetics of burst-like events in two different layers in the solar atmosphere.  Numerous authors, e.g. \cite{Ayres1995}, have shown flux--flux relationships between transition region lines and the X-ray flux. This work was taken one step further by \cite{Bruner1988} and \cite{Doyle1996} who showed a linear relationship between the flux of transition region lines and coronal lines, and the total radiative losses; hence the Si {\sc iv} 1394 {\AA} line is a good proxy to estimate the radiative losses in the atmosphere.

Seven out of the twelve datasets were already binned with a pixel scale of 0.33$''$, whilst the pixel scale for the remaining observations were approximately 0.17$''$. In order to have consistent pixel scales in all datasets, all observations were binned to have a pixel scale of 0.33$''$ before the application of the detection algorithm. In Fig.~\ref{fig1}, we show an example of a dataset selected for analysis here. This dataset (Set 1 from Table~\ref{tab1}) sampled a sunspot within AR NOAA 11836 in sit-and-stare mode between 15:59:35 UT and 17:54:29 UT on 2013 August 31. Both 1330 {\AA} and 1400 {\AA} SJ images were recorded with a 12 s cadence and a pixel size of 0.17$''$. The exposure time for the SJI observations was 4 s. In panels (a) and (b), we plot the SJI 1400 {\AA} and 1330 {\AA} images of the AR sampled at 17:12:28 UT, respectively. Corresponding AIA 1600 {\AA} and 193 {\AA} images collected by the Solar Dynamics Observatory's Atmospheric Imaging Assembly \citep[SDO/AIA][]{aia} instrument are shown in panel (c) and panel (d), respectively. The white box over-laid on panels (c) and (d) outlines the FOV of the IRIS SJI observation. The SJI data in Fig \ref{fig1} clearly display the presence of the localised brightenings we aim to study here.

\subsection{Burst Detection Algorithm}

\subsubsection{Burst selection:}

Intensities in IRIS SJI data are stored in DN units, however, the exposure time of images may not be constant during the observation. In order to correct for this effect, intensity values were divided by the corresponding exposure times before these exposure normalised images were used for the detection of implusive brightening events. Event selection was carried out by scanning the light curve of each pixel in the FOV. The algorithm starts by calculating the background intensity in each pixel from its time-series data before obtaining the standard deviation ($\sigma$) with respect to the background intensity. The algorithm first determines the minimum intensity of the pixel from its light curve and then calculates the average intensity of the light curve by excluding regions which are 5 times larger than the minimum intensity. This condition helps to remove impulsive brightening while calculating the background intensity. Impulsive bursts were then identified through the presence of intensity peaks which are significantly above background fluctuations for a user-defined time. We analysed the effects of varying the local intensity threshold between $3\sigma$ and $7\sigma$ as well as varying the minimum lifetime of bursts with these results being presented in the later sections of this article. The minimum burst lifetime studied was limited to three times the cadence in order to differentiate between bursts and cosmic ray spikes, which typically do not appear at the same location in subsequent images.

We also applied a condition that the enhanced emission identified as a bursts should be at least twice the average intensity of the entire FOV. A single pixel can have multiple bursts since there is a possibility of having multiple intensity maxima at the same location through time. As there could be some non-impulsive brightening events in the FOV, an impulsivity criterion was applied to remove sustained brightening events. For this, we calculated the average enhancement in intensity during the burst and divided this value by the background flux for that specific pixel. This ratio should be greater than 1 to be considered as an impulsive event. As no algorithm is perfect in detecting bursts, there is a possibility of excluding actual burst-like events as well as including non-burst regions. Choosing the correct combinations of parameters and conditions can minimise the mis-identification of the bursts.

\subsubsection{Area and lifetime of the burst:} 
Once an impulsive burst was identified in a pixel, an iterative approach was then used to calculate the total area and lifetime of the event. In the first step, the algorithm searched for connected pixels with intensity above the user defined intensity threshold during the peak time. Minimum area thresholding was applied within the algorithm to reduce incorrect identifications, with limits of $>$1, $>$2, or $>$3 pixels all studied. In the next step, the  algorithm searched for the impulsive events connected to the pixel in the previous time-step. The bf algorithm continued this iteration until it reached a time where there were no pixels above the intensity threshold. This same iterative method was also applied forward in time. Using this method, we were able to determine the starting time as well as the ending time of the burst. In this approach, the total area of the burst is a combination of all the spatially connected pixels throughout its lifetime. Once a burst location and time was identified, we removed these data points from the data cube to avoid multiple detections. 

\subsubsection{Energy determination:}
\begin{table*}
\caption{Power-law indices of the impulsive events detected in each dataset by defining bursts as peaks with intensity enhancement 5$\sigma$ and 3$\sigma$ above the background for $\sim 110$ s (second to fifth columns) and $\sim 60$ s (sixth to ninth columns). 'N/A' indicates too few events were returned to provide an accurate estimate of the power-law index.}

\centering
\begin{tabular}{|c|c|@{} c|@{} c|@{} c|@{} c|@{} c|@{} c|@{} c|}
\hline
Data & Power-law ($\gamma$)  & Power-law ($\gamma$) & Power-law ($\gamma$)   & Power-law ($\gamma$) & Power-law ($\gamma$)   & Power-law ($\gamma$) & Power-law ($\gamma$)   & Power-law ($\gamma$) \\
      & SJI 1400     & SJI 1330 &  SJI 1400     & SJI 1330 &  SJI 1400     & SJI 1330 &  SJI 1400     & SJI 1330\\
\hline
& \multicolumn{4}{c}{110 s} & \multicolumn{4}{c}{60 s}   \\
\hline
 & \multicolumn{2}{c}{5$\sigma$ above the background} & \multicolumn{2}{c}{3$\sigma$ above the background} & \multicolumn{2}{c}{5$\sigma$ above the background} & \multicolumn{2}{c}{3$\sigma$ above the background} \\
    \hline
Set 1  &$1.56 \pm{0.08}$ & $1.56 \pm{0.09}$ &$1.59 \pm{0.05}$ & $1.62 \pm{0.08}$ &$1.54 \pm{0.05}$  & $1.68 \pm{0.08}$      &$1.81 \pm{0.04}$  & $1.66 \pm{0.04}$  \\
Set 2  &$1.62 \pm{0.04}$ & $1.36 \pm{0.04}$ &$1.72 \pm{0.03}$ & $1.57 \pm{0.04}$  &$1.74 \pm{0.02}$  & $1.59 \pm{0.04}$      &$1.96 \pm{0.02}$  & $1.81 \pm{0.02}$   \\
Set 3 &$1.71 \pm{0.04}$ & $1.55 \pm{0.06}$   &$1.89 \pm{0.03}$ & $1.74 \pm{0.05}$  &$2.01 \pm{0.03}$  & $1.71 \pm{0.04}$      &$2.03 \pm{0.02}$  & $1.99 \pm{0.03}$  \\
Set 4 &$1.85 \pm{0.21}$ & $1.77 \pm{0.25}$    &$2.08 \pm{0.09}$ & $2.04 \pm{0.12}$  &$2.07 \pm{0.13}$  & $2.10 \pm{0.15}$      &$2.27 \pm{0.04}$  & $2.21 \pm{0.05}$ \\
Set 5  &$1.79 \pm{0.14}$ & $1.51 \pm{0.21}$   &$2.03 \pm{0.10}$ & $1.90 \pm{0.11}$ &$1.90 \pm{0.10}$  & $1.93 \pm{0.16}$      &$2.20 \pm{0.06}$  & $2.25 \pm{0.08}$  \\
Set 6 & $N/A$ & - & $N/A$ & - & $N/A$ & - &$1.62 \pm {0.20}$ & - \\
Set 7 & N/A & - & N/A & - &$1.62 \pm{0.20}$ & - &$1.75 \pm{0.31}$ & - \\
Set 8 &$1.83 \pm{0.02}$ & - &$1.83 \pm{0.03} $& - &$1.90 \pm{0.02} $& - & $1.91 \pm{0.02}$ & - \\
Set 9 &$1.93 \pm{0.03} $& - &$1.82 \pm{0.02} $& - &$1.94 \pm{0.03} $& - &$1.94 \pm {0.02} $& - \\
Set 10 &$1.63 \pm{0.07}$ & - &$1.76 \pm{0.04} $& - &$1.78 \pm{0.04} $& - &$1.98 \pm{0.02}$ & - \\
Set 11 &$1.89 \pm{0.16} $& - &$2.06 \pm{0.12} $& - &$1.97 \pm{0.07} $& - &$2.03 \pm{0.05} $& - \\
Set 12 & N/A & - &$1.36 \pm{0.15}$ & - & N/A & - &$1.95 \pm{0.13} $& - \\
\hline
\end{tabular}

\label{tab2}
\end{table*}

\begin{figure*}
\includegraphics[width=1.0\textwidth]{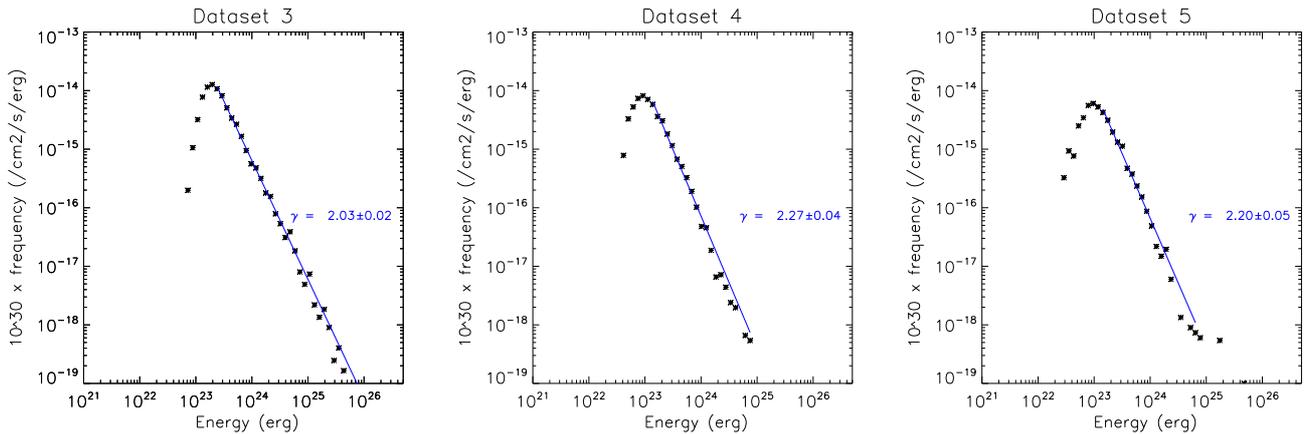}
\caption{Power-law distributions of impulsive events detected in the SJI 1400 channel with intensity enhancements above $3\sigma$, minimum areas of $>$3 pixels and minimum lifetimes of $60$ s for Datasets 3, 4, \& 5.}
\label{fig3}
\end{figure*}

 Before calculating the energy of these impulsive events it is necessary to express the SJI intensities in radiometric units, which provides the energy flux at the Sun. This can be done using the equation:
\begin{equation}
Flux\,(erg\,cm^{-2}\,s^{-1})=F\,(DN/s)\frac{4\pi\,E_{\lambda}\,k}{A\,\Omega}
\end{equation}
where $E_{\lambda}$ is the energy of the photon and $k$ is the factor that converts the DN to the number of photons. This factor is the ratio between gain and yield. The gain is the number of electrons released in the detector that yield 1 DN and the yield is the number of electrons released by one incident photon. {$\Omega$} is the SJI pixel size in steradian. $A$ is the post-launch effective area calculated using the `iris\_get\_response.pro' routine, which accounts for the degradation of the instrument since launch. The time over which the energy was calculated included the rise and decay phases of the bursts, delineated by the times when the gradient of the pixel intensity was zero before and after the burst. The energy radiated during the event can be calculated by integrating the radiated energy flux in time, in all bright pixels of each event. The background emission was subtracted while calculating the energy of the burst. 

\subsection{Testing the detection algorithm}

 Before performing the detection on real SJ images, we tested the detection algorithm on a synthetic dataset to verify its robustness. Firstly, we created a data cube that has base intensities which follow a Gaussian distribution. Since real solar data is not perfectly Gaussian, we introduced asymmetry by multiplying each pixel by an error function. Secondly, we scanned through the light curve of each pixel to obtain the average background intensity and the background fluctuations. Thirdly, we randomly selected 100 locations in the data cube using the IDL \textit{randomu} function and increased the intensity values in the corresponding locations by $10\, \sigma$ for a specified lifetime. These enhanced intensity pixels represents the observed bursts. Finally, we detected bursts with a minimum area of 3 pixels that are 3 $\sigma$ above the background and obtained the integrated intensity enhancement. 

The algorithm was successful in detecting simulated bursts; detecting 98 bursts out of 100. The code failed to detect 2 bursts since they were overlapping. In Fig.~\ref{fig_test}, we compare the intensity enhancement integrated over the actual area and lifetime of the bursts with the one obtained from the detected bursts. The integrated intensity enhancement represents the energy of the burst, which determines the power-law slope. Histograms for the actual values of integrated intensity enhancement are shown in the top panel of Fig~\ref{fig_test} while the value determined from the detected bursts is presented in the bottom panel. The similarity between these two plots implies how successful is the detection algorithm in determining the energies of the bursts.

\section{Results}
\label{Results}

Initially, we tested the effects of the chosen intensity threshold and minimum area conditions on the returned power-law indices. We present results from Dataset 1 in Fig.~\ref{fig2}, however, results from all other datasets are included in Table~\ref{tab2}. Some datasets do not have the SJI 1330 {\AA} channel, also for the short duration datasets we have too few points to provide accurate estimates of the power-law index when using a minimum lifetime condition of 110 s. In the top row of Fig \ref{fig2}, we plot radiative energy against frequency for burst events detected in the SJI 1400 \AA\ channel with a minimum lifetime of 110 s, with the three panels representing varying intensity thresholds. Similar results are plotted for the IRIS SJI 1330 {\AA} channel in the bottom panels. The intensity threshold used in the algorithm are shown on the top of each panel, with the distribution of bursts detected with enhancements of 3$\sigma$, 5$\sigma$, 7$\sigma$ above the background being shown in the left, middle, and right columns, respectively. The colour of the distributions represent the minimum area condition applied while detecting impulsive events with areas of $>$1, $>$2 and $>$3 pixels being plotted in black, red, and blue, respectively. Intuitively, more events are detected (especially at lower energies) when lower intensity thresholds are applied. Notably, despite fewer events being recorded, the maximum energies returned from the SJI 1330 \AA\ channel are larger than those returned from the SJI 1400 \AA\ channel by up to an order of magnitude with some features having energies $>10^{26}$ erg (see, for example, the middle panels of Fig. \ref{fig2}).

\begin{figure*}
\includegraphics[width=1.0\textwidth]{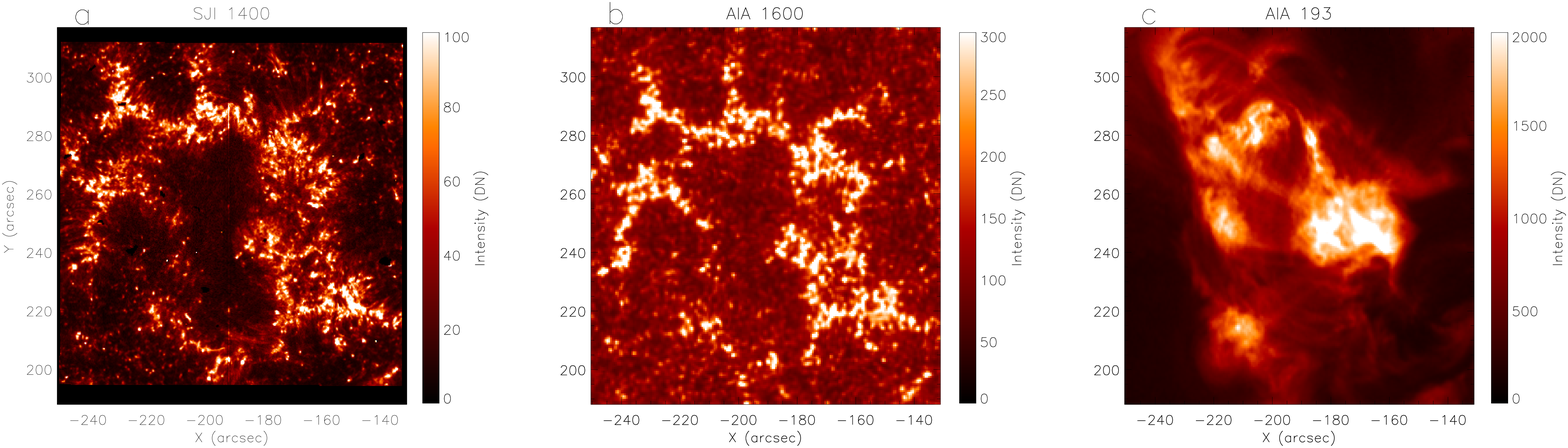}
\includegraphics[width=1.0\textwidth]{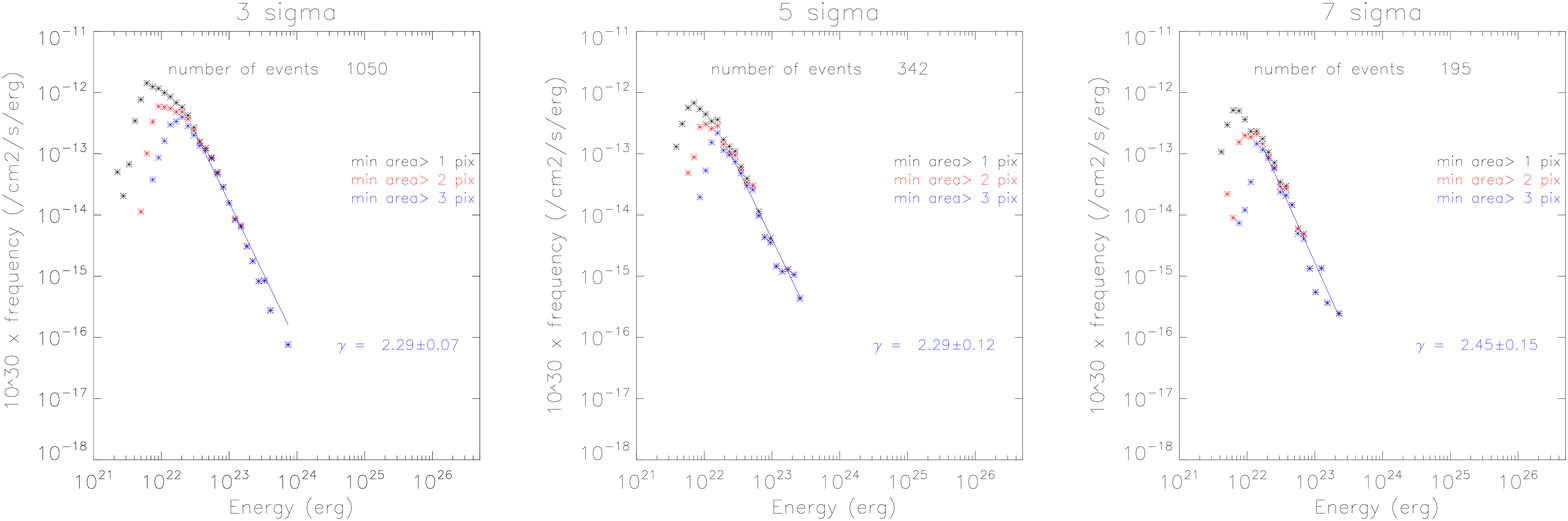}
\includegraphics[width=1.0\textwidth]{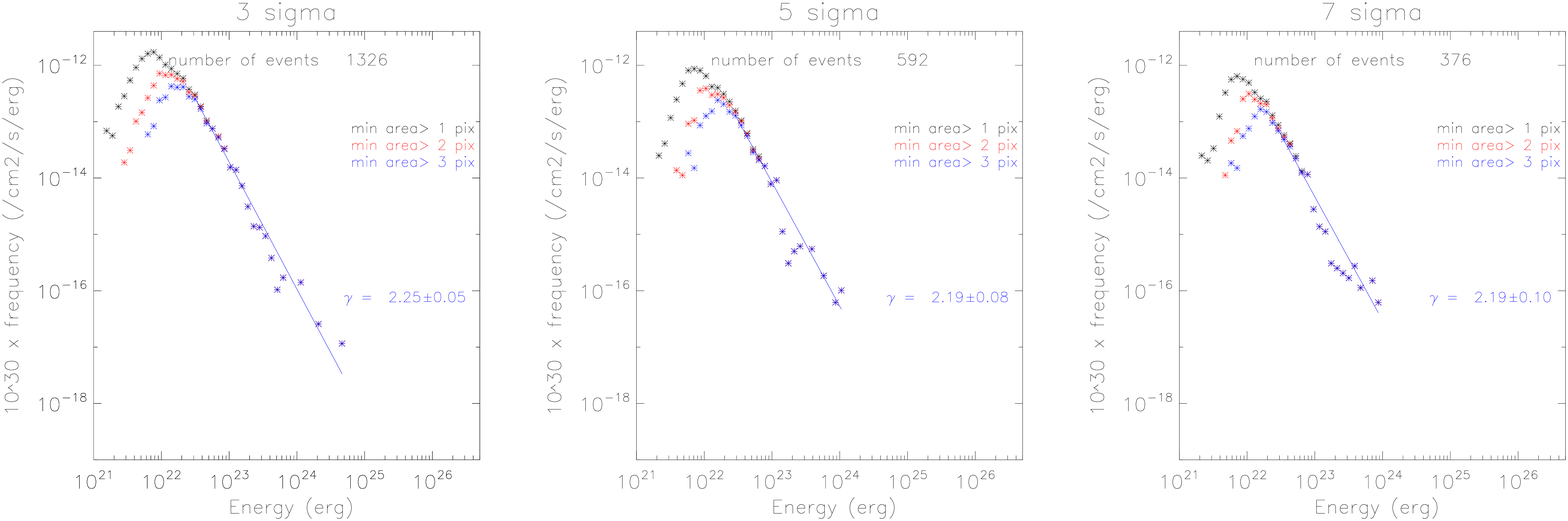}
\caption{Results of applying the detection algorithm for a plage region (Datasets 6 and 7) for varying minimum lifetime thresholds. The top row plots SJI 1400 {\AA} (left panel), AIA 1600 {\AA} (middle panel), and AIA 193 {\AA} (right panel) context images displaying the clear plage structuring in this FOV. The middle row plots energy distributions of impulsive events detected using a minimum lifetime condition of $\sim15$ s (three time-steps) from Dataset 6. The left, middle, and right panels display the distributions of impulsive events detected using $3\sigma$, $5\sigma$, and $7\sigma$ detection limits, respectively. The colours of the distribution depict the minimum area condition as in Fig.~\ref{fig2}. The bottom row plots the energy distributions of the impulsive events detected using a minimum lifetime condition of $\sim9$ s from Dataset 7. Again, the left, middle, and right panels display the impulsive events detected using $3\sigma$, $5\sigma$, and $7\sigma$ detection limits, respectively.}
\label{fig4}
\end{figure*}

Here, the radiative energies of the detected impulsive events range from $10^{22}$ erg to $10^{26}$ erg, thus these events could be considered to be nano-flares. At lower energies ($<10^{23}$ erg), the distribution of bursts fails to follow a power-law as the frequency of the detected bursts decreases with decreasing energy. The drop-off from the power law distribution at low energies is larger in bursts detected with less stringent intensity thresholds (black curves). This indicates that the drop off is likely to be caused by observational limitations (e.g., spatial and temporal resolutions) as it becomes increasingly difficult to accurately determine the parameters of bursts that are near the detection limit of the instrument. It should be noted, however, that changes made in the distributions with varying minimum area conditions can only be seen in the lower energies with the blue curve being completely superimposed on the red and black curves at higher energies ($>10^{23}$ erg).

\begin{table*}
\caption{Total radiative power obtained for all of the datasets using an intensity threshold of $3\sigma$ and a minimum lifetime threshold of $3$ times the cadence. $P_{tot}$ in the seventh column represents the total radiative power in the observed minimum and maximum energy range (see Eqn.~\ref{eqn2}). In the 8th column $P_{tot}$ is calculated for the energy range $10^{21}$ erg to observed maximum energy.}
\centering
\begin{tabular}{ |c|c|c|c|c|c|c|c|c}
\hline
Data  & Date of      & $E_{Min}$   & $E_{Max}$ & Power-law   & Minimum &$P_{tot}$  &$P_{tot}$ using \\
      & Observation  &  (erg)    &   (erg)           & index ($\gamma$)   &Lifetime (s)   &($erg~cm^{-2}~s^{-1}$) &$E_{Min}$=$10^{21}$  \\
\hline
Set 1 & 2013/08/31    &$7.86\times10^{22}$  & $3.27\times10^{25}$      &$1.93 \pm{0.04}$  & $\sim{36} $  &$1.72 \times 10^6$  & $2.72\times 10^{6}$ \\ 
Set 2 & 2013/12/17    &$9.33\times10^{22}$  & $6.46\times10^{26}$      &$1.96 \pm{0.02}$  & $\sim{60}  $ &$2.24 \times 10^6$  & $3.66\times 10^{6}$   \\
Set 3 & 2015/04/16    &$1.41\times10^{23}$  & $2.58\times10^{26}$      &$2.03 \pm{0.02}$  & $\sim{60}  $ &$2.47 \times 10^6$  & $4.67 \times 10^{6}$ \\
Set 4 & 2015/11/12    &$2.02\times10^{23}$ & $1.75\times10^{25}$       &$2.33 \pm{0.04}$  & $\sim{39}  $ &$1.06 \times 10^6$  & $5.40 \times 10^{6}$ \\
Set 5  & 2015/11/12   &$3.38\times10^{23}$  & $1.22\times10^{26}$    &$2.20 \pm{0.05}$  & $\sim{60} $    &$6.74 \times 10^5 $&$ 2.35 \times 10^{6}$ \\
Set 6  & 2017/03/26   &$7.05\times10^{22}$  & $1.64\times10^{24}$     &$2.29 \pm{0.07}$  & $\sim{15} $ &$2.13 \times 10^5$ &$7.53 \times 10^{5}$ \\
Set 7  & 2017/03/26   &$7.41\times10^{22}$  & $1.04\times10^{25}$     &$2.25 \pm{0.05}$  & $\sim{9} $ &$3.45 \times 10^5$ &$9.46 \times 10^{5}$ \\
Set 8  & 2017/05/01   &$1.47\times10^{23}$  & $7.17\times10^{25}$     &$2.08 \pm{0.01} $ & $\sim{36} $ &$6.17 \times 10^5$ &$1.10 \times 10^{6}$ \\
Set 9  & 2016/06/06   &$2.61\times10^{23}$  & $1.35\times10^{26}$     &$2.03 \pm{0.02}$  & $\sim{36} $ &$8.22 \times 10^5$ &$1.54 \times 10^{6}$ \\
Set 10  & 2014/01/07   &$1.94\times10^{23}$  & $7.60\times10^{25}$     &$1.98 \pm{0.02}$  & $\sim{60} $ &$3.15 \times 10^6$ & $6.01\times 10^6$\\
Set 11  & 2016/01/14   &$5.88\times10^{22}$  & $1.80\times10^{25}$     &$2.07 \pm{0.03} $ & $\sim{33} $ &$1.28 \times 10^6$ &$2.77 \times 10^{6}$\\
Set 12  & 2016/01/19   &$7.76\times10^{22}$  & $4.35\times10^{24}$     &$2.25 \pm{0.07}$  & $\sim{33}$  &$9.70 \times 10^5$ &$4.54 \times 10^{6}$\\

\hline
\end{tabular}
\label{tab3}
\end{table*}

As the transition region of ARs are generally very dynamic in nature, we note that if we reduce the intensity threshold too far it is likely that fluctuations not associated with impulsive features will be counted as bursts. The above effect must be considered while calculating $\gamma$ for the distribution from each dataset, meaning the power-law index was only calculated for energies above where the tail-off in frequencies are detected and for the most stringent area thresholding conditions (i.e., impulsive events with minimum area $>$3 pixels). For each of the intensity thresholds, the distribution displayed a clear linear behaviour on a logarithmic scale, with the resulting best fit and power-law index over-laid on each plot. Although the total number of events detected is reduced, as expected, with increasing sigma values, the $\gamma$-value remained stable within errors varying from $1.59\pm0.05$ ($3\sigma$) to $1.56\pm0.08$ ($7\sigma$) for the IRIS SJI $1400$ {\AA} channel and from $1.62\pm0.08$ ($3\sigma$) to $1.45\pm0.11$ ($7\sigma$) for the $1330$ {\AA} channel. This consistency gives us confidence in our results at higher energies. 

In Table~\ref{tab2} we display how the power-law index varies when impulsive events are identified with intensity enhancements of $5\sigma$ and $3\sigma$ above the background and lifetimes of over either 110 s (second to fifth columns) or 60 s (sixth to ninth columns) for both the SJI 1400 \AA\ and 1330 \AA\ channels. For the $1400$ \AA\ channel with a minimum lifetime of 60 s, $4$ ($2$) out of $12$ ($10$) datasets have power-law indices above $2$ for $3\sigma$ ($5\sigma$). With a minimum lifetime of 110 s this decreases to $3$ ($0$) out of $10$ ($9$) datasets . This result suggests that higher cadence data may return greater values of $\gamma$ in general. For the $1330$ \AA\ channel (where only $5$ datasets are available) these values are $1$ ($0$) out of $5$ for a minimum lifetime of $110$ s and $2$ ($1$) out of $5$ for a minimum lifetime of $60$ s. To display the accuracy of the power-law fitting, we plot three further examples (Datasets 3, 4, and 5) from the SJI 1400 \AA\ line at the $3\sigma$ level with a minimum lifetime of 60 s and a minimum area of $3$ pixels in Fig.~\ref{fig3}.

\begin{figure}
\includegraphics[width=0.48\textwidth]{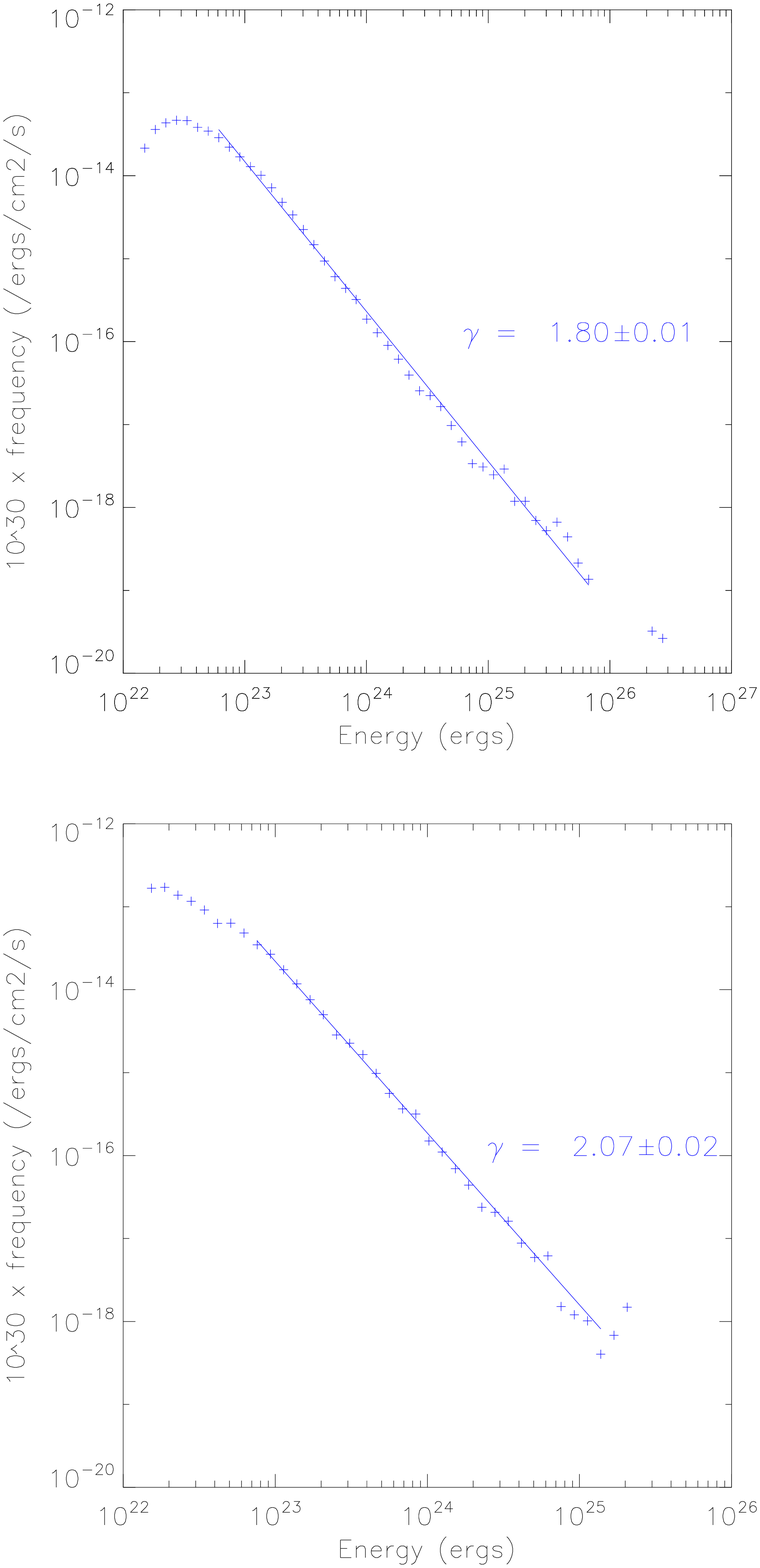}
\caption{Combined energy distribution of all 12 datasets using a cadence of $\sim60$ s (top panel). Combined energy distribution from all datasets studied with a cadence of less than $\sim35$ s, i.e., for Datasets 1, 4, 6, 7, 8, 9, 11 \& 12 (bottom panel).}
\label{fig5}
\end{figure}

Our results also indicate that the type of solar structures within the selected region (e.g., sunspot or plage) influences the power-law index. For regions dominated by sunspots, we always find that $\gamma <2$ for all threshold values when minimum lifetimes of 110 s and 60 s are considered; however, when the FOV is dominated by plage regions, the $\gamma$ values are higher, with $3$ out of $8$ completely plage datasets (see Tables~\ref{tab1} and \ref{tab2}) having $\gamma > $2 in the SJI $1400$ \AA\ channel at the $3\sigma$ level for both 110 s and 60 s cadences. We present results obtained from one plage region in more detail in Fig.~\ref{fig4}. In the top row of Fig.~\ref{fig4}, we plot an overview of the FOV as sampled by the SJI $1400$ \AA\ channel (left panel), the AIA $1600$ \AA\ channel (middle panel), and the AIA $193$ \AA\ channel (right panel). Large amounts of plage are evident but no sunspots or pores are present. Having examined the effects of the intensity threshold, minimum area, and solar structures within the FOV, we now move on to further investigate how the minimum lifetime criterion influences the returned $\gamma$-values.

Two datasets which sampled the plage region plotted in the top row of Fig.~\ref{fig4} were studied here, with both datasets being sampled at different cadences (Datasets 6 with 5 s and Dataset 7 with 3 s). The analysis of multiple datasets sampling the same FOV with different cadences allows us to remove the influence of different regions from our analysis. These datasets did not contain SJI $1330$ \AA\ data. In the middle and bottom rows of Fig.~\ref{fig4}, we plot the power-law distributions of impulsive events returned for SJI 1400 \AA\ data from Datasets 6 (middle row) and 7 (bottom row) where the minimum lifetime condition is set to three times the cadence. For all sigma values, $\gamma>2$ for both datasets. Notably, the power-law indices obtained for these datasets with minimum lifetime conditions are significantly higher (above errors) than those calculated from the same datasets when a 110 s or 60 s minimum lifetime was applied. The power-law indices calculated for three times the cadence with intensity thresholding set at $3\sigma$ are included in the fifth row of Table~\ref{tab3} for the remaining datasets. All but three datasets return $\gamma >2$.

We now move on to estimating the total radiative energies from each dataset. The total power per unit area of each impulsive event from the SJI 1400 \AA\ emission, $P_{1400}$, can be calculated by integrating the events energy times the frequency of occurrence of events per unit area per unit time, $f(E)$. Since the frequency of occurrence of these impulsive events follows a power-law distribution, the power per unit area can be written as follows:
\begin{equation}
P_{1400}=\int_{E_{min}}^{E_{max}} {f(E)EdE}=\frac{f_{0}}{-\gamma + 2} E^{-\gamma +2}\Big|_{E_{min}}^{E_{max}}, \ 
\label{eqn2}
\end{equation}
where $f_0$ is the normalization factor which can be calculated from the y-intercept of the energy distribution. $E_{min}$ and $E_{max}$ represent the observed minimum and maximum energy of the impulsive events from any given dataset. 

In order to calculate the total energy of the bursts we must first estimate the Si {\sc iv} 1394 {\AA} emission. This can be done from the SJI 1400 \AA\ channel by multiplying the burst energy obtained using our algorithm by 0.6 (i.e., assuming that around of the $\sim60\%$ of the SJI 1400 \AA\ intensity is contributed by Si {\sc iv} 1394 {\AA} emission; \citealt{hansteen_2014}). Once the Si {\sc iv} 1394 {\AA} emission had been estimated, we calculated the total radiative power ($P_{tot}$) per unit area using the equation:
\begin{equation}
log(P_{tot})=2.69+1.08\times log(P_{Si\_IV}).
\end{equation}
This is based on work by \cite{Bruner1988} and \cite{Doyle1996} who produced emission measure distributions for different solar regions (active regions, coronal holes, sunspots, `quiescent' regions and flares) plus a range of late-type stars based on data from the Hubble Space Telescope's International Ultraviolet Explorer and found a linear correlation between the total radiative output and that from a single spectral line. The typical error is $\pm$40\% \citep{Bruner1988}. The details of this calculation and the total radiative power obtained for all datasets is shown in Table \ref{tab3}. The derived values are typically only 10 -- 25\% of what is required in thermal energy to sustain the corona. If we were to extend the power-law distributions down to an energy of $10^{21}$ erg (as given in the final column in Table \ref{tab3}), and assume parity between the radiative energies reported here and thermal energy released by the impulsive events, then we would have approximately 25 -- 50\% of the required energy input to account for coronal heating.

Finally, we attempt to provide a more general answer as to the power-law index of these impulsive bursts in solar Active Regions. In order to do this we combined all of the datasets studied here in two ways. The first method combined the results obtained from each dataset when applying a minimum event lifetime of 60 s. The second method only considered datasets which facilitated a minimum lifetime condition of less than approximately $35$ s. The intensity threshold was set to $3\sigma$ and the area threshold was set to $>3$ pixels for both of these methods. In the top panel of Fig.~\ref{fig5}, the derived distribution from the first method is plotted. The noise within the data is extremely low, however, the power-law index is well below 2. When only datasets which could allow us to probe events with sub-minute lifetimes were studied, the power-law index increased such that $\gamma>2$. This result indicates that the analysis of extremely high-cadence IRIS SJI 1400 \AA\ data in the future may provide further in-sights into the energy input into the outer atmosphere from small-scale impulsive bursts.



\section{Discussion \& Summary}
\label{Discussion}

In this article, we have studied the effects of minimum area conditions, temporal resolution, intensity thresholding, and structuring within a FOV on the returned power-law indices of distributions of impulsive brightening events in the transition regions of solar ARs. We have studied 12 IRIS SJI datasets (detailed in Table~\ref{tab1}) using the automated detection algorithm described in Sect.~\ref{Observations}. Our results indicate that the minimum area condition has a limited influence on the power-law index, with the main effect of decreasing the minimum area being that one is able to observe lower energy events along the same power-law (see Figs.~\ref{fig2} and \ref{fig4}). The structuring within the FOV has a large effect on the $\gamma$-value with regions dominated by sunspots having smaller power-law indices than regions dominated by plage. It is interesting to note that if we select an intensity threshold of 3$\sigma$ above the background for impulsive event detections, then $3$ out of $8$ plage datasets have power-law indices higher than 2 for both $60$ s and $110$ s lifetime thresholds (see Tab.~\ref{tab2}).

In order to better understand the effects of cadence on the power-law indices, we also applied the algorithm to all datasets with a minimum lifetime value of three time-steps. With this criteria, $9$ of the $12$ datasets returned $\gamma>$2 indicating that cadence is an important factor for power-law determination (see Table~\ref{tab3}). The effects of cadence on the returned power-law are best shown in Fig.~\ref{fig5}, where combining all datasets with a minimum lifetime of 60 s returns a lower $\gamma$-value than combining all datasets with a 35 s minimum lifetime. Using the power-law indices obtained with a minimum lifetime of three times the cadence, we calculated the total radiative energy of the observed impulsive events for each dataset. The estimated total radiative power obtained for these datasets is typically only 25\% of the thermal energy required to maintain the temperature of the corona (see Table~\ref{tab3}). Although this would not be sufficient to heat solar plage regions, it would constitute a major contribution. However, if the derived slopes extend to energies of $10^{21}$ erg then the radiative energy would be at parity with the thermal energy required for coronal heating. 

In all of the regions studied, instrument sensitivity has a major impact on the detections at lower energies. \cite{Mondal2020} recently used the Murchison Widefield Array to observe the quiet Sun at radio frequencies. They detected impulsive emission with flux densities of about two orders of magnitude lower than the earlier attempts (a few mSFU where a Solar Flux Unit is $10^{-19}$ erg s$^{-1}$ cm$^{-2}$ Hz$^{-1}$). These impulsive events have duration of $\sim$1s and are present throughout the quiet solar corona. The estimate of the energy which must be dumped in the corona to generate these impulsive events is consistent with that required for coronal heating. In the events studied by these authors, the power-law indices were all $>$2. Combining this with our analysis implies that high cadence is an essential requirement for future analyses. In the future, the highest cadence IRIS SJI datasets available should be analysed in order to better understand the presence of impulsive brightenings at the smallest energy scales.

The Spectral Imaging of the Coronal Environment (SPICE) instrument on-board Solar Orbiter will observe a range of ultraviolet lines, the strongest being O {\sc vi} 1031.9 \AA\ with an estimated 8000 photons pixel$^{-1}$ s$^{-1}$ in an active region \citep{Anderson2019}. This would allow confirmation that brightening events from plage regions can account for the heating of the corona. Another up-coming facility is the Daniel K. Inouye Solar Telescope (DKIST) which will have a spatial resolution a factor of three better than currently available in the lower solar atmosphere. In the photosphere/lower chromosphere, \cite{reid2016} analysed EBs observed in H$\alpha$ line profiles recorded by the Swedish Solar Telescope \citep{Schramer2003} using automated methods. These observation also took place in ARs, therefore, it is possible to compare their statistics with the impulsive brightenings detected in the upper atmosphere of the AR. The energies of the EBs ranged from $10^{23}$ erg to $10^{26}$ erg. The power-law index obtained for energies of the EBs was 1.67. \cite{Nelson2013} also looked at brightenings in the H$\alpha$ line with the Dunn Solar Telescope but reduced the required intensity enhancement thereby detecting many more impulsive events. The radiative energy covered was from $2 \times 10^{22}$ to $4 \times 10^{25}$ erg with a slope of 2.14. Both of these power-law indices are comparable to those found here for a range of datasets. With the improved spatial resolution offered by DKIST, we will be able to resolve very small scale flaring events.

\section*{Acknowledgments}

Armagh Observatory and Planetarium is core funded by the Northern Ireland Executive through the Dept for Communities.
IRIS is a NASA small explorer mission developed and operated by LMSAL with mission operations executed at NASA
Ames Research center and major contributions to downlink communications funded by ESA and the
Norwegian Space Centre. We would like to thank the AIA and IRIS teams for providing valuable data. CJN and MM acknowledge support
from STFC under grant No. ST/P000304/1. We would like to thank Dr Aaron Reid for  discussions on the topic of Ellerman Bombs.

\section*{Data availability}
The datasets were derived from sources in the public domain: [IRIS; https://iris.lmsal.com/ , AIA; http://jsoc.stanford.edu/].

\bibliographystyle{mnras}
\bibliography{ref}

\end{document}